\documentstyle[12pt,epsf,epsfig]{article}
\newcount\timecount
\newcount\hours \newcount\minutes  \newcount\temp \newcount\pmhours
\hours = \time
\divide\hours by 60
\temp = \hours
\multiply\temp by 60
\minutes = \time
\advance\minutes by -\temp
\def\hour{\the\hours}
\def\minute{\ifnum\minutes<10 0\the\minutes
            \else\the\minutes\fi}
\def\clock{
\ifnum\hours=0 12:\minute\ AM
\else\ifnum\hours<12 \hour:\minute\ AM
      \else\ifnum\hours=12 12:\minute\ PM
            \else\ifnum\hours>12
                 \pmhours=\hours
                 \advance\pmhours by -12
                 \the\pmhours:\minute\ PM
                 \fi
            \fi
      \fi
\fi
}

\def\monthname{\relax\ifcase\month 0/\or January\or February\or
   March\or April\or May\or June\or July\or August\or September\or
   October\or November\or December\else\number\month/\fi}

\def\bold#1{\setbox0=\hbox{$#1$}%
     \kern-.025em\copy0\kern-\wd0
     \kern.05em\copy0\kern-\wd0
     \kern-.025em\raise.0433em\box0 }

\def\beq{\begin{equation}}
\def\eeq{\end{equation}}

\def\ga{\mathrel{\raise.3ex\hbox{$>$\kern-.75em\lower1ex\hbox{$\sim$}}}}
\def\la{\mathrel{\raise.3ex\hbox{$<$\kern-.75em\lower1ex\hbox{$\sim$}}}}
\def\gev{{\rm \, Ge\kern-0.125em V}}
\def\tev{{\rm \, Te\kern-0.125em V}}
\def\gyr{{\rm \, G\kern-0.125em yr}}

\def\gappeq{\mathrel{\rlap {\raise.5ex\hbox{$>$}}
{\lower.5ex\hbox{$\sim$}}}}
\def\lappeq{\mathrel{\rlap{\raise.5ex\hbox{$<$}}
{\lower.5ex\hbox{$\sim$}}}}
\def\Toprel#1\over#2{\mathrel{\mathop{#2}\limits^{#1}}}

\def\m12{m_{1\!/2}}

\begin{document}
\begin{titlepage}
\pagestyle{empty}
\baselineskip=21pt
\rightline{\tt hep-ph/0303043}
\rightline{CERN--TH/2003-051}
\rightline{UMN--TH--2132/03}
\rightline{TPI--MINN--03/07}
\vskip 0.2in
\begin{center}
{\large {\bf Supersymmetric Dark Matter in Light of WMAP}} \\
\end{center}
\begin{center}
\vskip 0.2in
{\bf John~Ellis}$^1$, {\bf Keith~A.~Olive}$^{2}$, {\bf Yudi Santoso}$^{2}$ 
and {\bf Vassilis~C. Spanos}$^{2}$
\vskip 0.1in
{\it
$^1${TH Division, CERN, Geneva, Switzerland}\\
$^2${William I. Fine Theoretical Physics Institute, \\
University of Minnesota, Minneapolis, MN 55455, USA}}\\
\vskip 0.2in
{\bf Abstract}
\end{center}
\baselineskip=18pt \noindent

We re-examine the parameter space of the constrained minimal
supersymmetric extension of the Standard Model (CMSSM), taking account of
the restricted range of $\Omega_{CDM} h^2$ consistent with the WMAP data.
This provides a significantly reduced upper limit on the mass of the
lightest supersymmetric particle LSP: $m_\chi \lappeq 500$~GeV for $\tan
\beta \lappeq 45$ and $\mu > 0$, or $\tan \beta \lappeq 30$ and $\mu < 0$,
thereby improving the prospects for measuring supersymmetry at the LHC,
and increasing the likelihood that a 1-TeV linear $e^+ e^-$ collider would
be able to measure the properties of some supersymmetric particles.

\vfill
\leftline{CERN--TH/2003-051}
\leftline{March 2003}
\end{titlepage}
\baselineskip=18pt

\section{Introduction}

The recent data from the WMAP satellite~\cite{wmap} confirm with greater
accuracy the standard cosmological model, according to which the current
energy density of the Universe is comprised by about 73~\% of dark energy
and 27~\% of matter, most of which is in the form of non-baryonic dark
matter. The WMAP data further tell us that very little of this dark matter
can be hot neutrino dark matter, and the reported re-ionization of the
Universe when the redshift $z \sim 20$ is evidence against warm dark
matter. WMAP quotes a total matter density $\Omega_m h^2 =
0.135^{+0.008}_{-0.009}$ and a baryon density $\Omega_b h^2 = 0.0224 \pm
0.0009$~\cite{wmap}, from which we infer the following 2-$\sigma$ range
for the density of cold dark matter: $\Omega_{CDM} h^2 =
0.1126^{+0.0161}_{-0.0181}$. This range is consistent with that inferred
from earlier observations~\cite{cmb,MS}, but is significantly more
precise.

It has been appreciated for some time that the lightest supersymmetric
particle (LSP) is a suitable candidate for this non-baryonic cold dark
matter~\cite{EHNOS}. The LSP is stable in supersymmetric models where $R$
parity is conserved, and its relic density falls naturally within the
favoured range if it weighs less than $\sim 1$~TeV. This statement may be
made more precise in the minimal supersymmetric extension of the Standard
Model (MSSM), in which the LSP is expected to be the lightest neutralino
$\chi$, particularly if the soft supersymmetry-breaking mass terms
$m_{1/2}, m_0$ are constrained to be universal at an input GUT scale: the
constrained MSSM (CMSSM). Since $\Omega_\chi h^2 \propto m_\chi n_\chi$,
where $n_\chi$ is the relic LSP number density, and $n_\chi$ typically
increases as the universal soft supersymmetry-breaking mass parameters are
increased, one would expect the upper limit on $m_\chi$ to {\it decrease}
when the upper limit on $\Omega_{CDM} h^2$ is {\it decreased}. Compared
with taking this upper limit to be 0.3, as we and
others~\cite{EFGOSi,eos2,otherOmega} have done previously, taking
$\Omega_{CDM} h^2 < 0.129$ as suggested by the WMAP data - which is line
with estimates from previous CMB determinations used in~\cite{prev} - may
therefore be expected to improve significantly the corresponding upper
limit on $m_\chi$.

Such is indeed the case for $\tan \beta \la 45$ and $\mu > 0$, or for
$\tan \beta \la 30$ and $\mu < 0$, as we show below, where the largest
values of $m_\chi$ are found in the $\chi-{\tilde \tau}$ coannihilation
region. We find for these cases that $m_{1/2} \lappeq 900-1200$~GeV for
$\Omega_{CDM} h^2 <   0.129$,
whereas $m_{1/2} \lappeq 1400-1750$~GeV would have been allowed for
$\Omega_{CDM} h^2 < 0.3$. Correspondingly, the upper limit on the LSP  
mass
becomes $m_\chi \lappeq 400-500$~GeV, rather than $m_\chi \lappeq
600-700$~GeV   as found
previously~\cite{efo,moreco}.
This stronger upper limit improves the prospects for measuring
supersymmetry at the LHC. Also, it would put sleptons within reach of a
1-TeV linear $e^+ e^-$ linear collider, whereas previously a
centre-of-mass energy above 1.2~TeV might have appeared 
necessary~\cite{LC}. All  
the
above remarks would also apply if other particles also contribute to
$\Omega_{CDM}$.

For any fixed value of $\tan \beta$ and sign of $\mu$, only a narrow
region of CMSSM parameter space would be allowed if $0.094 < \Omega_\chi
h^2 < 0.129$, as would be implied by WMAP if there are no other  
significant
contributors to the density of cold dark matter. The previous `bulk'
regions of parameter space at small values of $m_0$ and $m_{1/2}$ now
become quite emaciated, and the previous coannihilation strips now  
become
much narrower, as do the rapid-annihilation funnels that appear at  
larger
$\tan \beta$~\cite{funnel,EFGOSi}.
However, unlike the coannihilation strips, the
rapid-annihilation funnels still extend to very large values of $m_0$  
and
$m_{1/2}$, the absolute upper limit on $m_\chi$ is much weaker for
$\tan \beta \gappeq 50$ if $\mu > 0$, or for $\tan \beta \gappeq 35$ if
$\mu < 0$~\footnote{Strictly speaking, there is also a filament of
parameter space extending to large $m_\chi$ in
the `focus-point' region~\cite{focus} at
large $m_0$, to which we return later.}. The narrowness of the
preferred region implies that $\tan \beta$ could in principle be
determined from measurements of $m_{1/2}$ and $m_0$, as we discuss  
later.

\section{WMAP Constraint on the CMSSM Parameter Space}

\begin{figure}
\begin{center}
\mbox{\epsfig{file=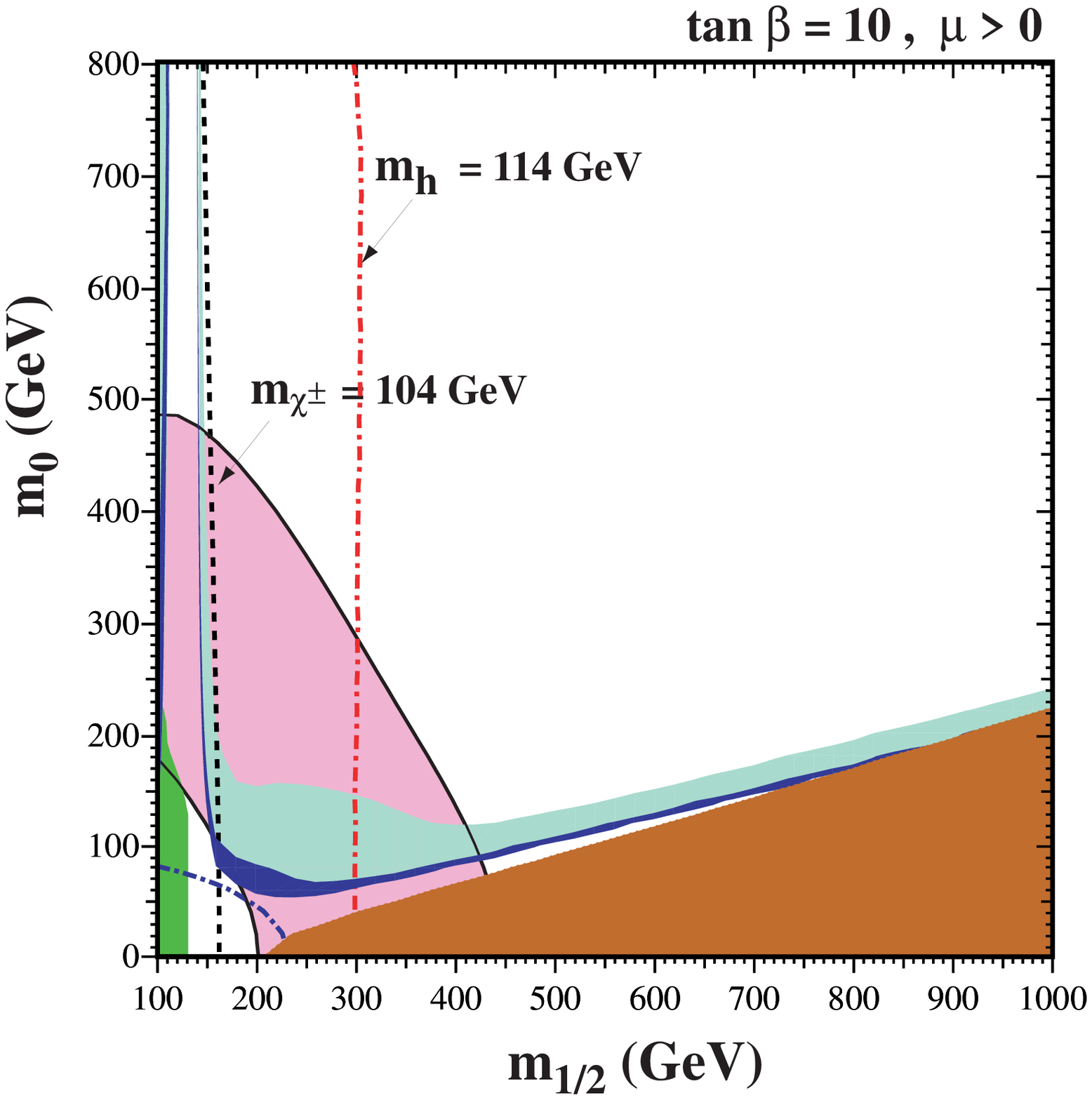,height=7cm}}
\mbox{\epsfig{file=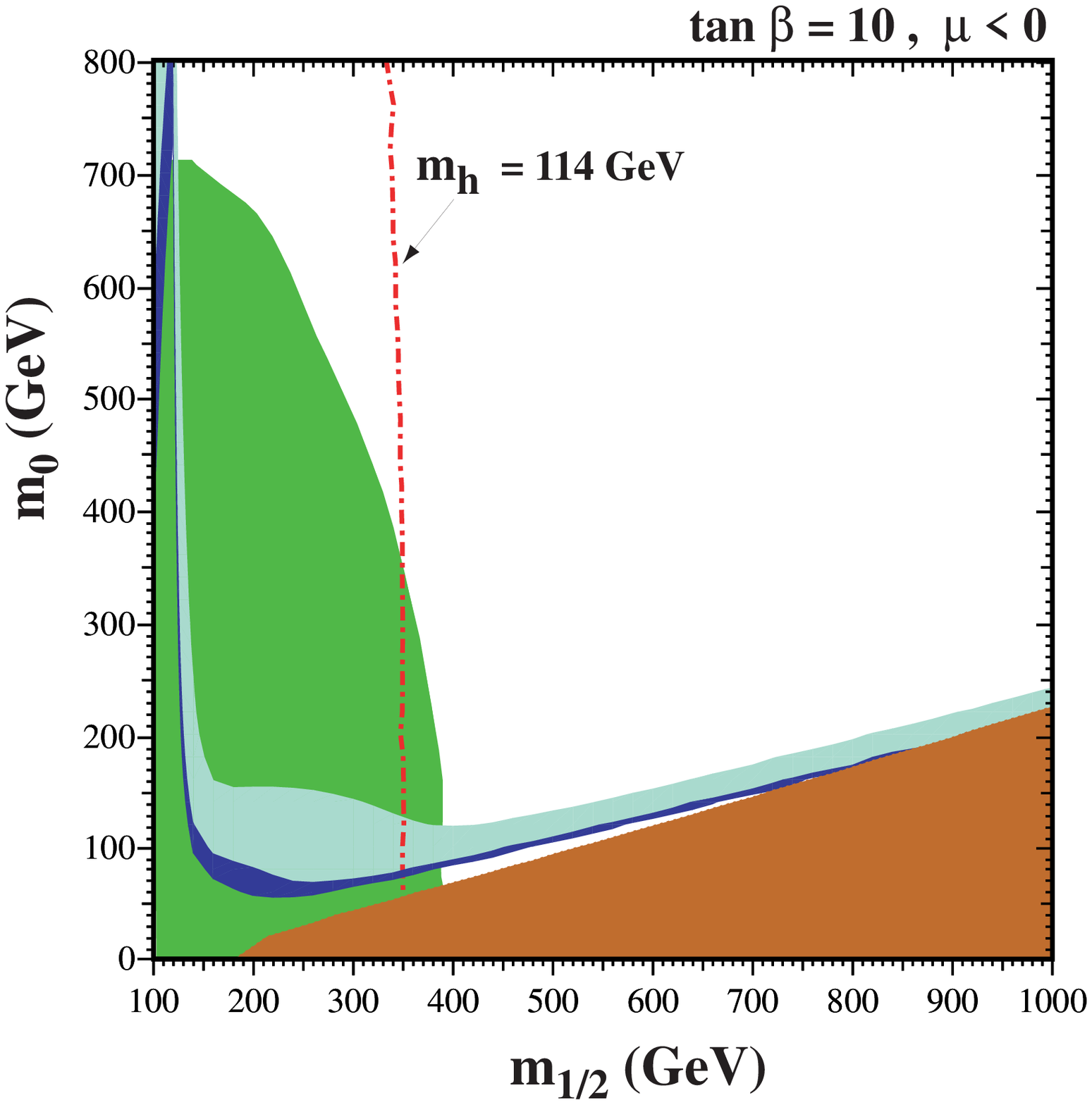,height=7cm}}
\end{center}
\begin{center}
\mbox{\epsfig{file=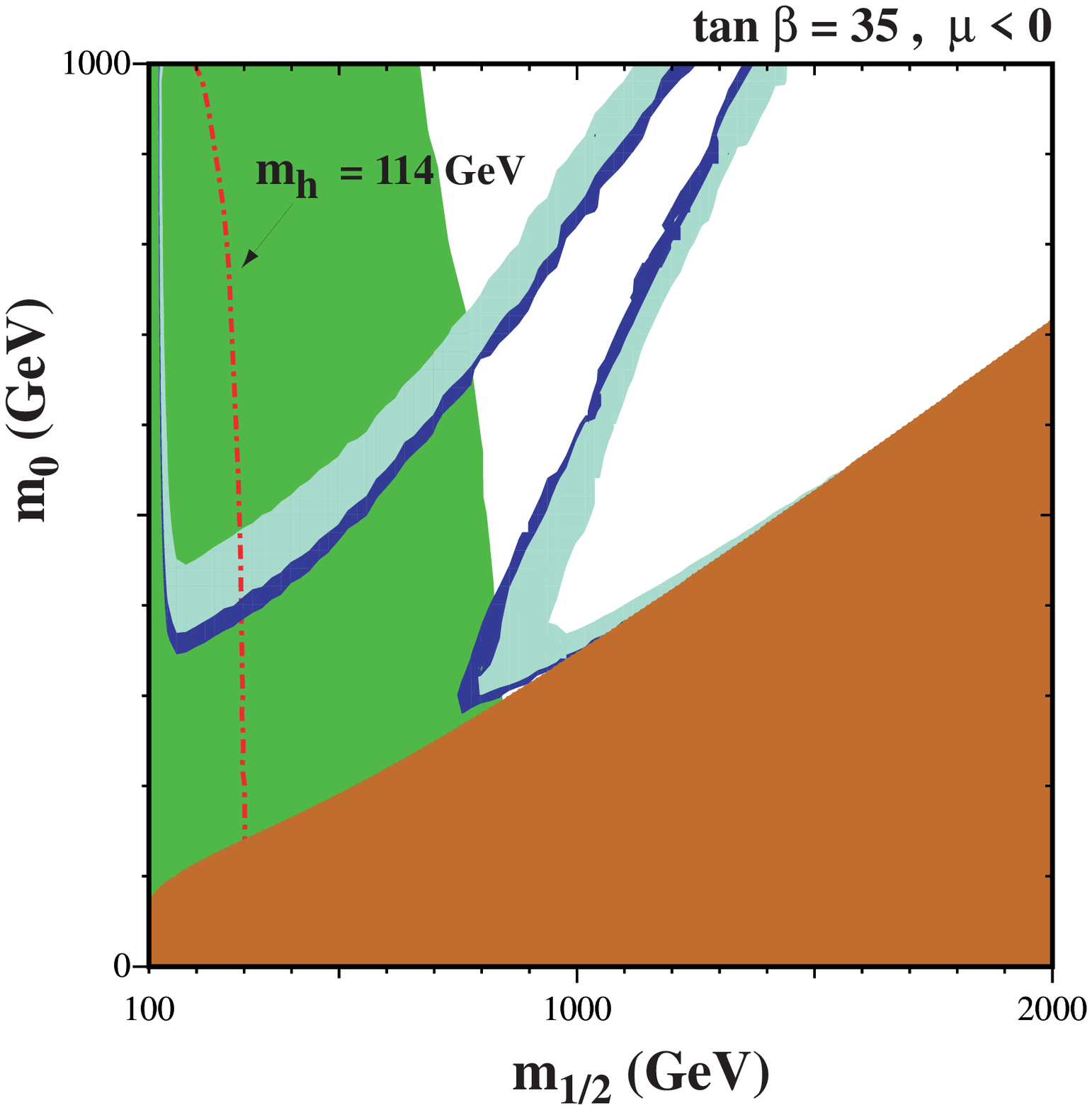,height=7cm}}
\mbox{\epsfig{file=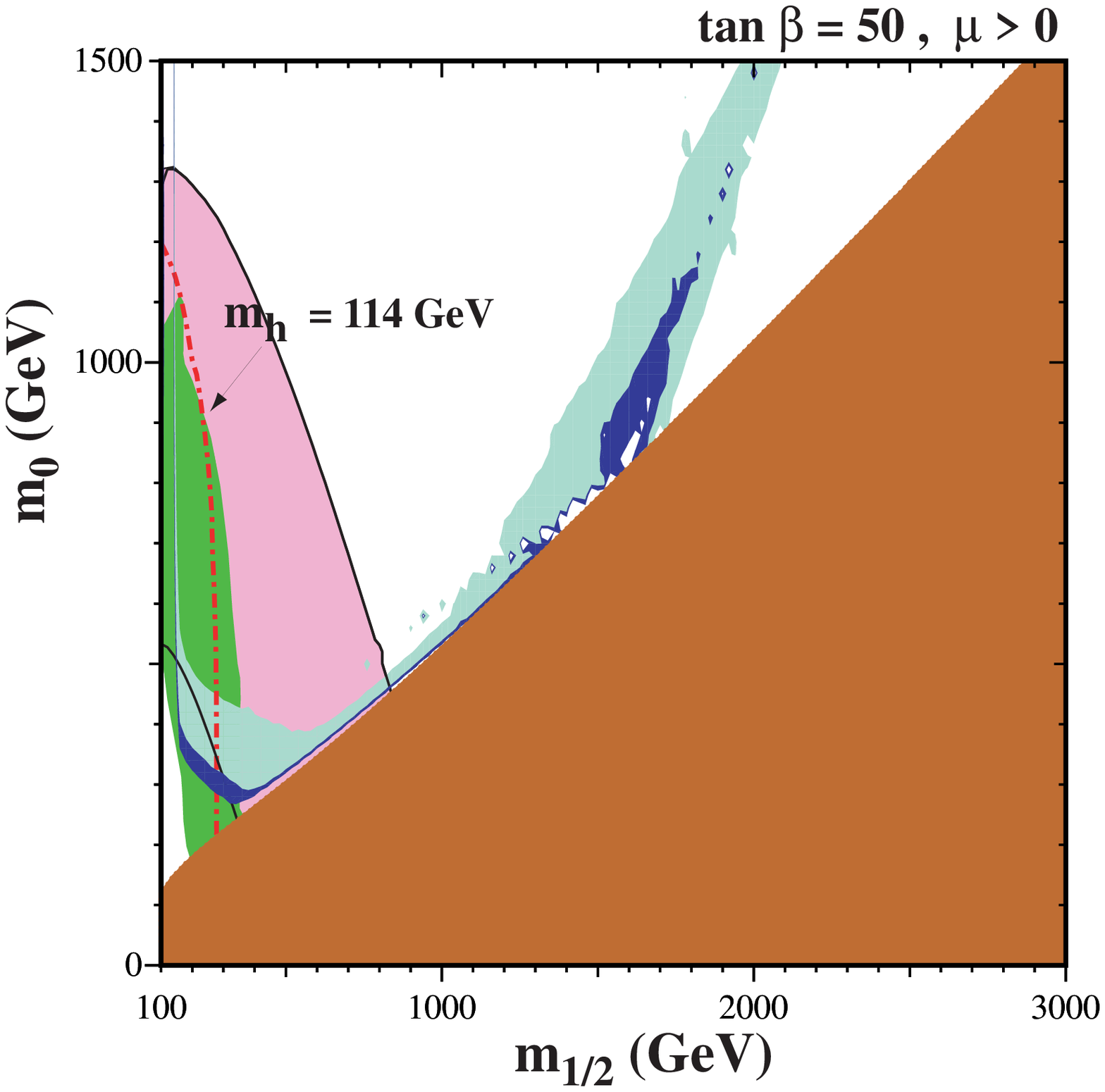,height=7cm}}
\end{center}
\caption{\label{fig:Omega}\it
The $(m_{1/2}, m_0)$ planes for (a) $\tan\beta = 10, \mu > 0$, (b)
$\tan\beta = 10, \mu < 0$, (c) $\tan\beta = 35, \mu < 0$, and (d)
$\tan\beta = 50, \mu > 0$. In each panel, the
region allowed by the older cosmological constraint $0.1 \le \Omega_\chi
h^2 \le 0.3$ has medium shading, and the region allowed by the newer
cosmological constraint $0.094 \le \Omega_\chi h^2 \le 0.129$ has very 
dark
shading. The disallowed region where $m_{\tilde \tau_1} < m_\chi$ has dark  
(red) shading.
The regions excluded by $b \rightarrow s \gamma$ have medium
(green) shading, and those in panels (a,d) that are favoured by $g_\mu - 
2$ at the 2-$\sigma$ 
level have medium (pink) shading. A dot-dashed line in panel (a)
delineates the LEP constraint on the $\tilde e$ mass and
the contours $m_{\chi^\pm} = 104$~GeV ($m_h = 114$~GeV) are shown
as near-vertical black dashed (red dot-dashed) lines in panel (a) (each 
panel).
}
\end{figure}

Fig.~\ref{fig:Omega} displays the allowed regions of the CMSSM parameter
space for (a) $\tan \beta = 10, \mu > 0$, (b) $\tan \beta = 10, \mu <  
0$,
(c) $\tan \beta = 35, \mu < 0$, and (d) $\tan \beta = 50, \mu > 0$. We
have taken $m_t = 175$~GeV and $A_0 = 0$ in all of the results shown 
below. In each panel, we
show the regions excluded by the LEP lower limits on
$m_{\tilde e}, m_{\chi^\pm}$ and $m_h$, as well as those ruled out by $b
\to s \gamma$ decay~\cite{bsg} as discussed in \cite{efoso}. In panels (a)
and (d) for
$\mu > 0$, we   also display
the regions favoured by the recent BNL measurement~\cite{newBNL}
of $g_\mu -2$ at the 2-$\sigma$ level,
relative to the calculation of the Standard Model based
on $e^+ e^-$ data at low energies~\cite{Davier}.

Also shown in Fig.~\ref{fig:Omega} are the `old' regions where $0.1 <
\Omega_\chi h^2 < 0.3$, and the `new' regions where $0.094 < \Omega_\chi
h^2 < 0.129$. We see immediately that (i) the cosmological regions are
generally much narrower, and (ii) the `bulk' regions at small $m_{1/2}$
and $m_0$ have almost disappeared, in particular when the laboratory
constraints are imposed. Looking more closely at the coannihilation
regions, we see that (iii) they are significantly truncated as well as
becoming much narrower, since the reduced upper bound on $\Omega_\chi h^2$
moves the tip where $m_\chi = m_{\tilde \tau}$ to smaller $m_{1/2}$. It is
this effect that provides the reduced upper bound on $m_\chi$ advertized
earlier. In panels (c) and (d), we see rapid-annihilation funnels that
(iv) are also narrower and extend to lower $m_{1/2}$ and $m_0$ than
previously. They weaken significantly the upper bound on $m_\chi$ for
$\tan \beta \gappeq 35$ for $\mu < 0$ and $\tan \beta \gappeq 50$ for $\mu
> 0$.

We take this opportunity to comment on some calculational details
concerning the rapid-annihilation funnels. Comparison with other studies
of these regions~\cite{otherOmega,LS} has shown the importance of treating
correctly the running mass of the bottom quark, as we have done in
previous works~\cite{EFGOSi}. Also important is the correct
treatment of annihilation rates across the convolution of two Boltzmann
distributions when the cross section varies rapidly, as in the
rapid-annihilation funnels, and we have taken the opportunity of this
paper to improve our previous treatment. The results on the low-$m_{1/2}$
sides of the the rapid-annihilation funnels are indistinguishable from
those shown previously, apart from the narrowing due to the smaller
allowed range of $\Omega_\chi h^2$. However, there are more significant
differences on the high-$m_{1/2}$ sides of the the rapid-annihilation
funnels, where our previous approximation was less adequate. This is most
noticeable for the case $\tan \beta = 50, \mu > 0$ shown in panel (d) of
Fig.~\ref{fig:Omega}, where the two strips with relic density in the
allowed range are all but merged. A second side of the rapid-annihilation
funnel becomes distinctly visible when $\tan \beta \ge 51$,
but the gap between the two sides is much narrower than we found
previously. This is also true for $\mu < 0$, and is exemplified in
Fig.~\ref{fig:Omega}c for $\tan \beta = 35$ where both sides of the
funnel region are clearly distinct\footnote{Note that the irregularities
seen in the cosmological regions are a result of the resolution used to
produce the figures.}.

Before discussing further our results, we comment on the potential
impact of a re-evaluation of $m_t$. A recent re-analysis by the D0
collaboration favours a central value some 5~GeV higher than the central
value $m_t = 175$~GeV that we use~\cite{new_mtop}. If confirmed, this
would shift the displayed contours of $m_h = 114$~GeV to lower $m_{1/2}$,
e.g., from $m_{1/2} \simeq 300$~GeV to $m_{1/2} \simeq 235$~GeV in the
case $\tan \beta = 10, \mu > 0$. This would not affect the upper bound on
$m_\chi$ that we quote later, but it would weaken the lower limit on
$m_\chi$ in cases where this is provided by the LEP Higgs limit.

The focus-point region would, however, be affected spectacularly by any
such increase in $m_t$, shifting to much larger $m_0$. In the analysis of
\cite{focus}, for $\tan\beta = 10$ and $m_{1/2} = 300$ GeV, the
focus-point is pushed up from $m_0 \simeq 2200$ GeV to $m_0 \simeq 4200$
GeV when $m_t$ is increased from 175 GeV to 180 GeV, and for $\tan
\beta = 50$, it is pushed up from $m_0 \simeq 1800$ GeV to 
$m_0 \simeq 3000$ GeV. In our treatment of the CMSSM, we in fact find no
focus-point region when $m_t = 180$ GeV. In view of this instability  in
the fixed-point region, we do not include it in our subsequent analysis:
our limits should be understood as not applying to this region, though we
do note that it would also be further narrowed by the more restricted
range of
$\Omega_\chi h^2$.

\begin{figure}
\begin{center}
\mbox{\epsfig{file=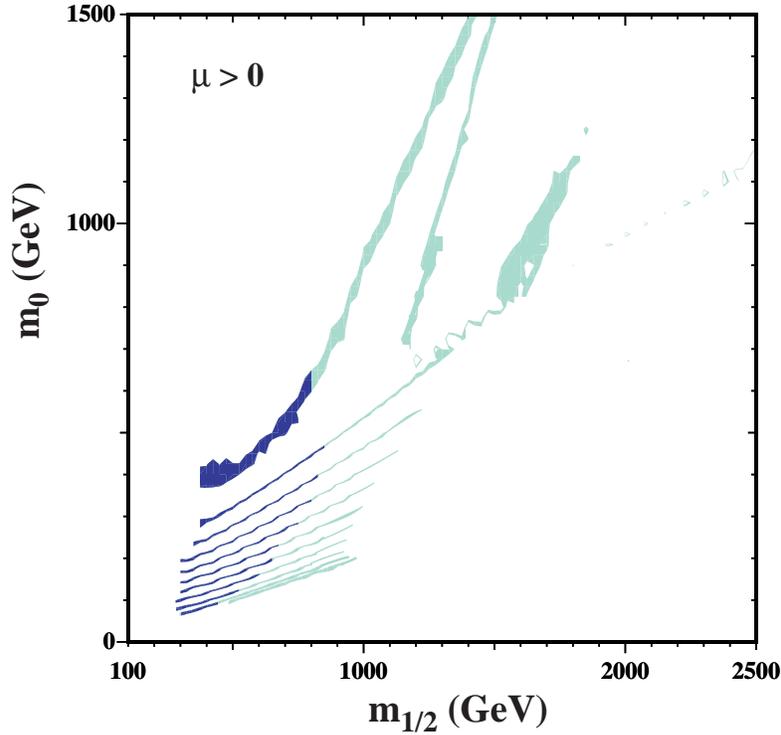,height=10cm}}
\end{center}
\caption{\label{fig:strips}\it
The strips display the regions of the $(m_{1/2}, m_0)$ plane that are
compatible with $0.094 < \Omega_\chi h^2 < 0.129$ and the laboratory
constraints for $\mu > 0$ and $\tan \beta = 5, 10, 15, 20, 25, 30,
35, 40, 45, 50, 55$. The parts of the strips compatible with $g_\mu - 2$ 
at the 2-$\sigma$ level have darker shading.
}
\end{figure}

We display in Fig.~\ref{fig:strips} the strips of the $(m_{1/2}, m_0)$
plane allowed by the new cosmological constraint $0.094 < \Omega_\chi h^2
< 0.129$ and the laboratory constraints listed above, for $\mu > 0$ and
values of $\tan \beta$ from 5 to 55, in steps $\Delta ( \tan \beta ) = 5$.
We notice immediately that the strips are considerably narrower than the
spacing between them, though any intermediate point in the $(m_{1/2},
m_0)$ plane would be compatible with some intermediate value of $\tan
\beta$. The right (left) ends of the strips correspond to the maximal
(minimal) allowed values of $m_{1/2}$ and hence $m_\chi$\footnote{The
droplets in the upper right of the figure are due to coannihilations
when $\tilde \tau$ is sitting on the Higgs pole. Here this occurs at
$\tan \beta = 45$.}. The lower bounds on $m_{1/2}$ are due to the Higgs 
mass constraint for $\tan \beta \le 23$, but are determined by the $b \to 
s \gamma$ constraint for higher values of $\tan \beta$. The
upper bound on $m_{1/2}$ for $\tan \beta
\gappeq 50$ is clearly weaker, because of the rapid-annihilation regions.

Also shown in Fig.~\ref{fig:strips} in darker shading are the restricted
parts of the strips that are compatible with the BNL measurement of $g_\mu
- 2$ at the 2-$\sigma$ level, if low-energy $e^+ e^-$ data are used to
calculate the Standard Model contribution~\cite{Davier}. If this
constraint is imposed, the range of $m_{1/2}$ is much reduced for any
fixed value of $\tan \beta$, and in particular the upper bound on
$m_{1/2}$ is significantly reduced, particularly for $\tan \beta \gappeq
50$. However, there is in general no change in the lower bound on
$m_{1/2}$.

\section{Improved Upper Limit on the LSP Mass}

\begin{figure}
\begin{center}
\mbox{\epsfig{file=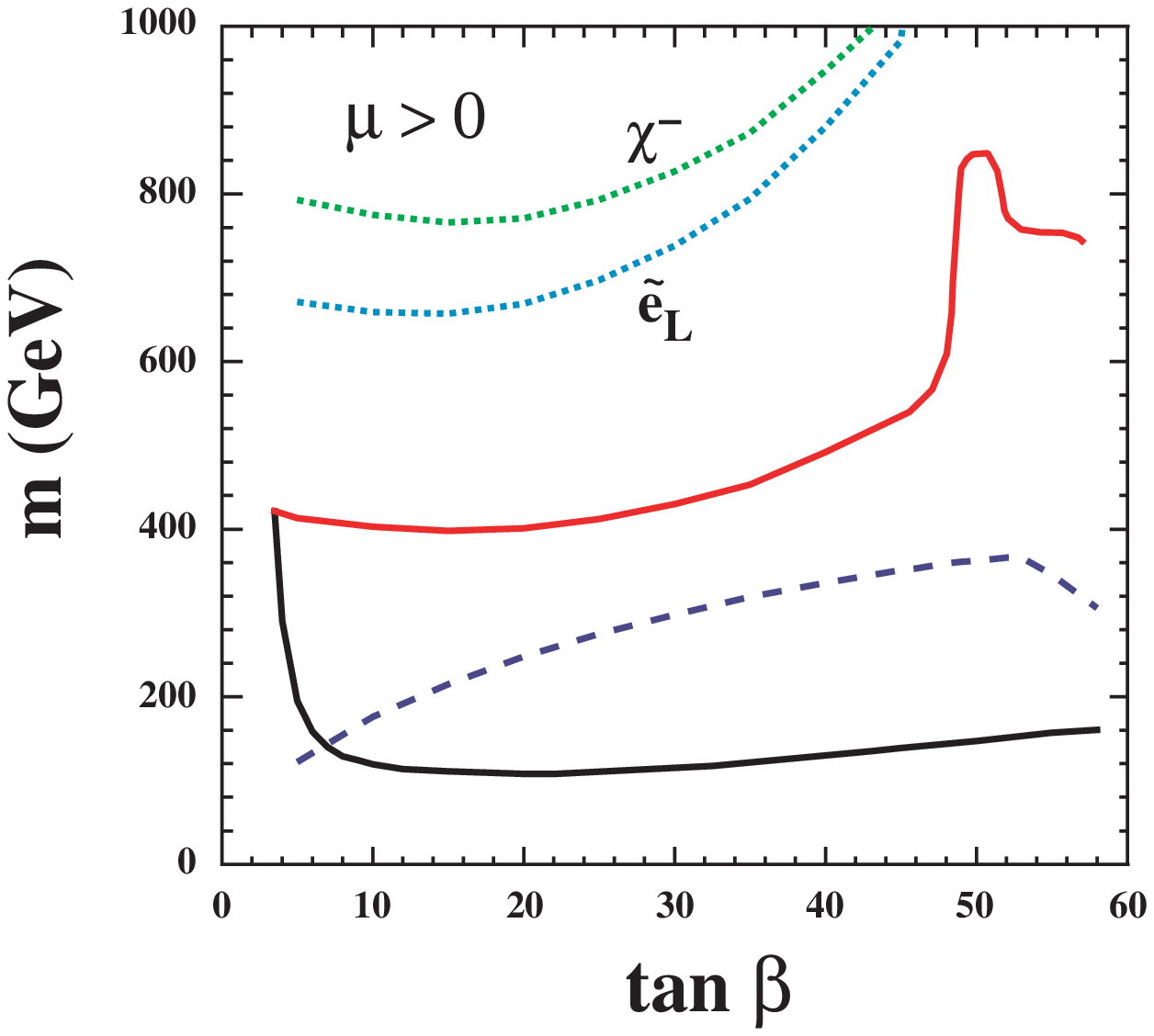,height=7cm}}
\mbox{\epsfig{file=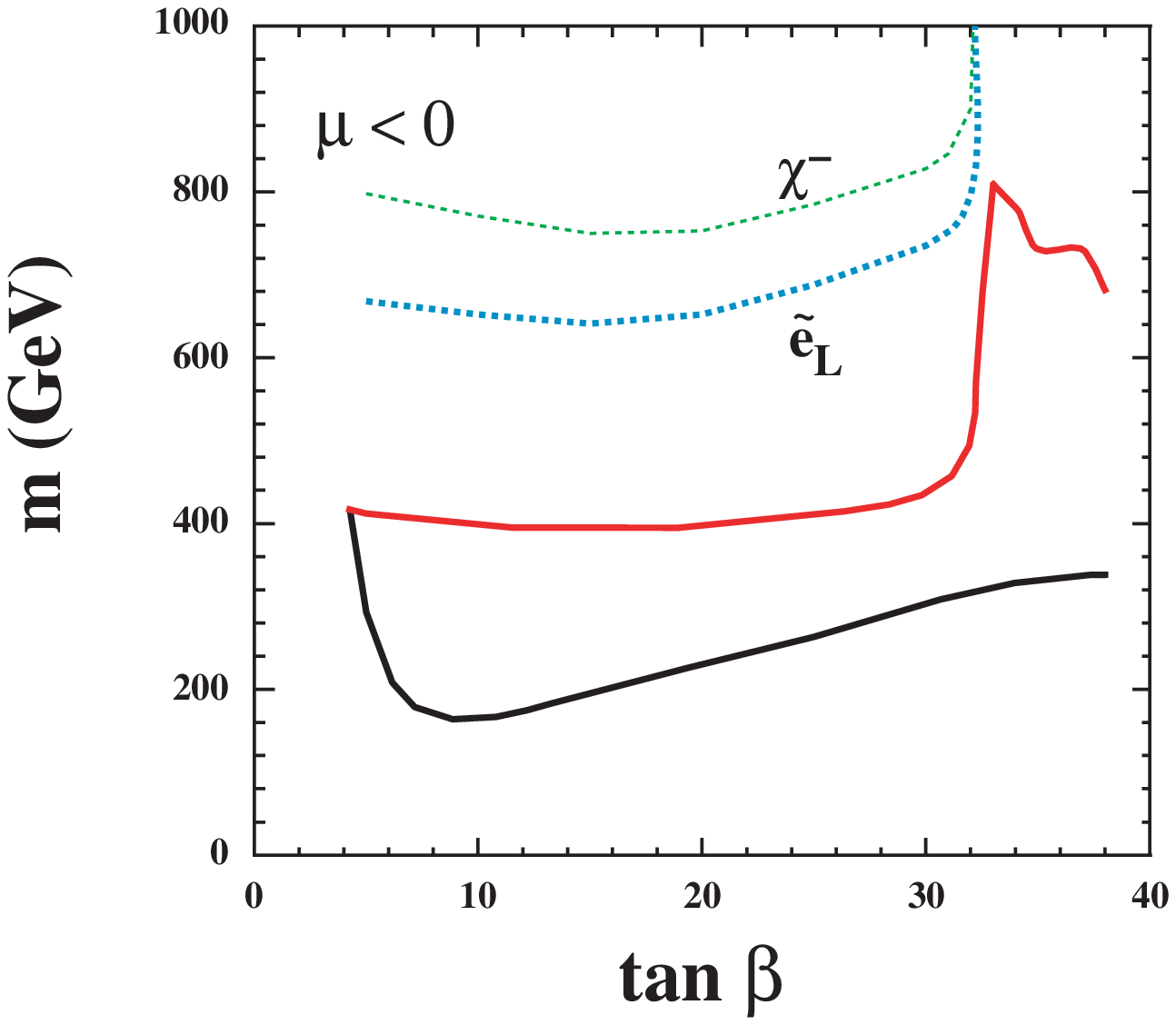,height=7cm}}
\end{center}
\caption{\it
The ranges of $m_\chi$ allowed by cosmology and other constraints, for
(a) $\mu > 0$ and (b) $\mu < 0$. Upper limits without (red solid line) 
and with (blue dashed line) the $g_\mu - 2$ constraint are shown for $\mu 
> 0$: the lower limits are shown as black solid lines. Note the 
sharp increases in the upper limits for $\tan \beta \gappeq 50, \mu > 0$ 
and $\tan \beta \gappeq 35, \mu < 0$ due to the rapid-annihilation 
funnels. Also shown as dotted lines are the ${\tilde e_L}$ and $\chi^\pm$ 
masses at the tips of the coannihilation tails.} 
\label{fig:ranges}
\end{figure}

We now draw some conclusions from Fig.~\ref{fig:strips}. Its implications
for the allowed range of the LSP mass $m_\chi$ as a function of $\tan
\beta$ are displayed in Fig.~\ref{fig:ranges}. As already mentioned, the
upper limit is rather weak for $\tan \beta \gappeq 50$ when $\mu > 0$. 
However, for $\tan \beta < 40$, we find the absolute upper bound
\beq
m_\chi \; \la \; 500~{\rm GeV},
\label{chibound}
\eeq
to be compared with the range up to $\simeq 650$~GeV that we found with
the old cosmological relic density constraint. Also shown in
Figs.~\ref{fig:strips} and \ref{fig:ranges} is the strengthened upper
bound on $m_{1/2}$ and $m_\chi$ that would apply if one used the $g_\mu -
2$ constraint. We find
\beq
m_\chi \; \la \; 370~{\rm GeV},
\label{g-2bound}
\eeq
for all values of $\tan \beta$. Fig.~\ref{fig:ranges}(a) also shows the  
lower bound on $m_\chi$ as a function of $\tan \beta$, leading to
\beq
m_\chi \; > \; 108~{\rm GeV},
\label{lowerbound}
\eeq
for all values of $\tan \beta$, with the minimum occurring around $\tan
\beta = 23$, when the $b \to s \gamma$ constraint begins to dominate over
the Higgs mass constraint. As such, this lower  limit
depends on the calculation of $m_h$, for which we use the  latest version
of {\tt FeynHiggs}~\cite{FeynHiggs}. This calculation has  an estimated
theoretical uncertainty $\sim 2$~GeV, and is very sensitive to 
$m_t$. The lower bound (\ref{lowerbound}) would become $> 86$~GeV (with
the minimum occurring at $\tan \beta \simeq 18$) if we  used $m_t \simeq
180$~GeV in {\tt FeynHiggs}, or $> 84$~GeV (with
the minimum occurring at $\tan \beta \simeq 17$) if we allowed for a 
$2$~GeV reduction in the calculated value for the nominal value of $m_t$.

We do not show the plot corresponding to Fig.~\ref{fig:strips} for $\mu <
0$, but we do show in Fig.~\ref{fig:ranges}(b) the corresponding lower and 
upper bounds on $m_\chi$ for $\tan \beta \lappeq 40$. We note again that 
the upper
bound would rise rapidly for larger $\tan \beta$, due to the appearance of
a rapid-annihilation funnel analogous to that appearing for $\tan \beta
\gappeq 50$ when $\mu > 0$. In the $\mu < 0$ case, there is no possibility
of compatibility with $g_\mu - 2$ when the $e^+ e^-$ data are used. We 
find the following range for $\tan \beta \le 30$:
\beq
160~{\rm GeV} < m_\chi < 430~{\rm GeV}
\label{negbound}
\eeq
for $\mu < 0$, with the lower bound being provided by $b \to s \gamma$
for $\tan \beta > 8$.

We note that the upper and lower limits meet when $\tan \beta = 3.5 (4.3)$
for 
$\mu > (<) 0$, implying that lower values of $\tan \beta$ are not allowed 
within our analysis of the CMSSM. This lower bound on $\tan \beta$ is 
strengthened to 7 if the $g_\mu - 2$ constraint is included.

\section{Implications for Supersymmetric Phenomenology}

The reduced upper limit on the sparticle mass scale
improves the prospects for measuring supersymmetry at the LHC. Also, the
fact that only a narrow strip in the $(m_{1/2}, m_0)$ plane is allowed for
each value of $\tan \beta$ offers the possibility of determining $\tan
\beta$ once $m_{1/2}$ and $m_0$ are known, e.g, from measurements at the
LHC. We have discussed previously the sensitivity of $\Omega_\chi h^2$ to
variations in the CMSSM parameters~\cite{ftuning}, and that analysis can
be adapted to the present situation.  Since the typical separation between
strips with $\Delta ( \tan \beta ) = 5$ is $\Delta m_0 \simeq 25$~GeV and
the width of a typical strip is $\Delta m_0 \simeq 5$~GeV, it would in
principle be possible to fix $\tan \beta$ with an accuracy $\Delta ( \tan
\beta ) \simeq 1$ using measurements of $m_{1/2}$ and $m_0$ alone for a
fixed value of $A_0$ (taken to be 0 here). The required accuracy in
$m_{1/2}$ is not very demanding, since the strips are nearly horizontal,
but $m_0$ would need to be determined with an accuracy
$\Delta m_0 \lappeq 5$~GeV. It is interesting to compare with the
accuracies in $m_0$, $m_{1/2}$ and $\tan \beta$ reported
in~\cite{Hinchliffe} for the case $m_0 = 100$~GeV, $m_{1/2} = 300$~GeV and
$\tan \beta = 2.1$. This value of $\tan \beta$ is not compatible with our
analysis, but the expected accuracies $\Delta ( m_0 ) \sim 10$~GeV and
$\Delta ( m_{1/2} ) = 20$~GeV would already allow interesting crosschecks
of the value of $\tan \beta$ extracted from a fit to the LHC data with
that required by cosmology.

The strengthened upper bound (\ref{chibound}) on $m_\chi$ may also have
important consequences for linear $e^+ e^-$ collider physics \cite{kamon}.
At the tip of the coannihilation region, which corresponds to the upper
bound in Fig.~\ref{fig:ranges}, we have $m_\chi = m_{\tilde \tau_1}$,
with the
${\tilde \mu_R}$ and ${\tilde e_R}$ not much heavier. Therefore, a linear
$e^+ e^-$ collider with centre-of-mass energy 1~TeV would be able to
produce these sleptons. This conclusion holds only if one restricts
attention to the CMSSM, as studied here, ignores the focus-point region as
also done here, and discards large values of $\tan \beta$. Moreover, we
note that the left-handed sleptons are somewhat heavier at the tip of the
CMSSM cosmological region: $m_{\tilde \ell_L} \gappeq 700$~GeV, as shown 
by the pale blue dotted lines in Fig.~\ref{fig:ranges}, and the lightest
chargino has $m_{\chi^\pm} \gappeq 800$~GeV (green dotted lines).  
However, the strengthened upper limit on $\Omega_\chi h^2$ that has been
provided by WMAP does strengthen the physics case for a TeV-scale
linear $e^+ e^-$ collider, compared with~\cite{LC}.

\section{Perspective}

In general, the WMAP constraint on $\Omega_\chi h^2$ put supersymmetric
phenomenology in a new perspective, essentially by reducing the
dimensionality of the parameter space: one can now consider $m_0$ to be
(almost) fixed in terms of the other parameters. The `Snowmass
lines'~\cite{lines} now intersect the allowed cosmological region in just
one (fuzzy) point each. The post-LEP benchmark points~\cite{benchmark}
have values of $\Omega_\chi h^2$ that lie above the WMAP range. However,
most of them can easily be adapted, in the `bulk' and coannihilation
regions simply by reducing $m_0$. An exception is benchmark point H, which
was chosen at the tip of a coannihilation tail: WMAP would require this to
be brought down to lower $m_{1/2}$, which would make it easier to detect
at the LHC or a future linear $e^+ e^-$ linear collider. A more detailed
update of the CMSSM benchmarks will be presented elsewhere.

\vskip 0.5in
\vbox{
\noindent{ {\bf Acknowledgments} } \\
\noindent We would like to thank H. Baer and A. Belyaev for helpful
discussions. The work of K.A.O., Y.S., and V.C.S. was supported in part
by DOE grant DE--FG02--94ER--40823.}


\begin{thebibliography}{99}

\bibitem{wmap}
C.~L.~Bennett {\it et al.},
arXiv:astro-ph/0302207;
D.~N.~Spergel {\it et al.},
arXiv:astro-ph/0302209.


\bibitem{cmb}
A.~T.~Lee {\it et al.} [MAXIMA-1 Collaboration],
Astrophys.\ J.\  {\bf 561} (2001) L1
[arXiv:astro-ph/0104459];
C.~B.~Netterfield {\it et al.}  [Boomerang Collaboration],
Astrophys.\ J.\  {\bf 571} (2002) 604
[arXiv:astro-ph/0104460];
C.~Pryke {\it et al.} [DASI Collaboration],
Astrophys.\ J.\  {\bf 568} (2002) 46
[arXiv:astro-ph/0104490].



\bibitem{MS}
A.~Melchiorri and J.~Silk,
Phys.\ Rev.\ D {\bf 66} (2002) 041301
[arXiv:astro-ph/0203200];
J.~L.~Sievers {\it et al.},
arXiv:astro-ph/0205387.



\bibitem{EHNOS}
J.~Ellis, J.~S.~Hagelin, D.~V.~Nanopoulos, K.~A.~Olive
and M.~Srednicki, Nucl. Phys. B {\bf 238} (1984) 453; see also
H.~Goldberg, Phys. Rev. Lett. {\bf 50} (1983) 1419.


\bibitem{EFGOSi}
J.~R.~Ellis, T.~Falk, G.~Ganis, K.~A.~Olive and M.~Srednicki,
  Phys.\ Lett. {\bf B510} (2001) 236
[arXiv:hep-ph/0102098].



\bibitem{eos2} J.~R.~Ellis, K.~A.~Olive and Y.~Santoso,
  New Jour.\ Phys.  {\bf 4} (2002) 32
[arXiv:hep-ph/0202110].

\bibitem{otherOmega}
L.~Roszkowski, R.~Ruiz de Austri and T.~Nihei,
JHEP {\bf 0108} (2001) 024
[arXiv:hep-ph/0106334];
A.~Djouadi, M.~Drees and J.~L.~Kneur,
JHEP {\bf 0108} (2001) 055
[arXiv:hep-ph/0107316];
H.~Baer, C.~Balazs and A.~Belyaev,
JHEP {\bf 0203} (2002) 042
[arXiv:hep-ph/0202076].


\bibitem{prev}
A.~B.~Lahanas, D.~V.~Nanopoulos and V.~C.~Spanos,
Phys.\ Rev.\ D {\bf 62} (2000) 023515
[arXiv:hep-ph/9909497];
Mod.\ Phys.\ Lett.\ A {\bf 16} (2001) 1229
[arXiv:hep-ph/0009065];
Phys.\ Lett. {\bf B518} (2001) 94
[arXiv:hep-ph/0107151];
V.~Barger and C.~Kao,
Phys.\ Lett. {\bf B518} (2001) 117
[arXiv:hep-ph/0106189];
R.~Arnowitt and B.~Dutta,
arXiv:hep-ph/0211417.

\bibitem{efo} J. Ellis, T. Falk, and K.A. Olive,  Phys.\ Lett. {\bf  
B444} (1998) 367
[arXiv:hep-ph/9810360];
J. Ellis, T. Falk, K.A. Olive and M. Srednicki,  Astr. Part. Phys.
{\bf 13} (2000) 181
[Erratum-ibid.\  {\bf 15} (2001) 413]
[arXiv:hep-ph/9905481].



\bibitem{moreco}
M.~E.~G\'omez, G.~Lazarides and C.~Pallis,
Phys. Rev. D {\bf D61} (2000) 123512
[arXiv:hep-ph/9907261];
  Phys.\ Lett. {\bf B487} (2000) 313
[arXiv:hep-ph/0004028];
  Nucl. Phys. B {\bf B638} (2002) 165
[arXiv:hep-ph/0203131];
R.~Arnowitt, B.~Dutta and Y.~Santoso,
  Nucl. Phys. B {\bf B606} (2001) 59
[arXiv:hep-ph/0102181].
T.~Nihei, L.~Roszkowski and R.~Ruiz de Austri,
  JHEP {\bf 0207} (2002) 024
[arXiv:hep-ph/0206266].

\bibitem{LC}
J.~R.~Ellis, G.~Ganis and K.~A.~Olive,
Phys.\ Lett.\ B {\bf 474} (2000) 314
[arXiv:hep-ph/9912324].

\bibitem{funnel}
M.~Drees and M.~M.~Nojiri,
Phys.\ Rev.\ D {\bf 47} (1993) 376
[arXiv:hep-ph/9207234];
H.~Baer and M.~Brhlik,
Phys.\ Rev.\ D {\bf 53} (1996) 597
[arXiv:hep-ph/9508321];
A.~B.~Lahanas, D.~V.~Nanopoulos and V.~C.~Spanos,
Phys.\ Rev.\ D {\bf 62} (2000) 023515
[arXiv:hep-ph/9909497];
H.~Baer, M.~Brhlik, M.~A.~Diaz, J.~Ferrandis, P.~Mercadante,  
P.~Quintana and X.~Tata,
Phys.\ Rev.\ D {\bf 63} (2001) 015007
[arXiv:hep-ph/0005027];
A.~B.~Lahanas, D.~V.~Nanopoulos and V.~C.~Spanos,
Mod.\ Phys.\ Lett.\ A {\bf 16} (2001) 1229
[arXiv:hep-ph/0009065].






\bibitem{focus}
J.~L.~Feng, K.~T.~Matchev and T.~Moroi,
  Phys.\ Rev.\ Lett.\  {\bf 84} (2000) 2322;
J.~L.~Feng, K.~T.~Matchev and T.~Moroi,
Phys.\ Rev. {\bf D61} (2000) 075005;
J.~L.~Feng, K.~T.~Matchev and F.~Wilczek,
Phys.\ Lett.   {\bf B482} (2000) 388.




\bibitem{bsg}
M.S. Alam et al., [CLEO Collaboration], Phys.\ Rev.\ Lett.\ {\bf 74}
(1995) 2885 as updated in
S.~Ahmed et al., {CLEO CONF 99-10};
BELLE Collaboration, BELLE-CONF-0003, contribution to the 30th
International conference on High-Energy Physics, Osaka, 2000.
See also
K.~Abe {\it et al.},  [Belle Collaboration],
[arXiv:hep-ex/0107065];
L.~Lista  [BaBar Collaboration],
[arXiv:hep-ex/0110010];
K.~Chetyrkin, M.~Misiak and M.~Munz,
Phys.\ Lett.\ B {\bf 400} (1997) 206
[Erratum-ibid.\ B {\bf 425}, 414 (1997)]
[hep-ph/9612313];
T.~Hurth,
hep-ph/0106050;
C. Degrassi, P. Gambino and G.~F. Giudice,
JHEP {\bf 0012} (2000) 009 [arXiv:hep-ph/0009337];
M.~Carena, D.~Garcia, U.~Nierste and C.~E.~Wagner,
Phys. Lett. B {\bf 499} (2001) 141
[arXiv:hep-ph/0010003].
P.~Gambino and M.~Misiak,
Nucl.\ Phys.\ B {\bf 611} (2001) 338;
D.~A.~Demir and K.~A.~Olive,
Phys.\ Rev.\ D {\bf 65} (2002) 034007
[arXiv:hep-ph/0107329].

\bibitem{efoso}
J.~R.~Ellis, T.~Falk, K.~A.~Olive and Y.~Santoso,
Nucl.\ Phys.\ B {\bf 652} (2003) 259
[arXiv:hep-ph/0210205].

\bibitem{newBNL}
G.~W.~Bennett {\it et al.}  [Muon $g-2$ Collaboration],
Phys.\ Rev.\ Lett.\  {\bf 89} (2002) 101804
[Erratum-ibid.\  {\bf 89} (2002) 129903]
[arXiv:hep-ex/0208001].


\bibitem{Davier}
M.~Davier, S.~Eidelman, A.~Hocker and Z.~Zhang,
decays:
arXiv:hep-ph/0208177; see also
K.~Hagiwara, A.~D.~Martin, D.~Nomura and T.~Teubner,
arXiv:hep-ph/0209187;
F.~Jegerlehner, unpublished, as reported in
M.~Krawczyk,
Acta Phys.\ Polon.\ B {\bf 33} (2002) 2621
[arXiv:hep-ph/0208076].



\bibitem{LS}
A.~B.~Lahanas and V.~C.~Spanos,
Eur.\ Phys.\ J.\ C {\bf 23} (2002) 185
[arXiv:hep-ph/0106345].


\bibitem{new_mtop}
The RunII D0 top group, see \\
{\tt http://www-d0.fnal.gov/Run2Physics/top/conf.html}.

\bibitem{FeynHiggs}
S.~Heinemeyer, W.~Hollik and G.~Weiglein,
Comput.\ Phys.\ Commun.\  {\bf 124} (2000) 76
[arXiv:hep-ph/9812320];
S.~Heinemeyer, W.~Hollik and G.~Weiglein,
Eur.\ Phys.\ J.\ C {\bf 9} (1999) 343
[arXiv:hep-ph/9812472].

\bibitem{ftuning}
J.~R.~Ellis and K.~A.~Olive,
Phys.\ Lett.\ B {\bf 514} (2001) 114
[arXiv:hep-ph/0105004].

\bibitem{Hinchliffe}
I.~Hinchliffe, F.~E.~Paige, M.~D.~Shapiro, J.~Soderqvist and W.~Yao,
Phys.\ Rev.\ D {\bf 55} (1997) 5520   
[arXiv:hep-ph/9610544].

\bibitem{kamon}
T.~Kamon, R.~Arnowitt, B.~Dutta and V.~Khotilovich,
arXiv:hep-ph/0302249.

\bibitem{lines}
B.~C.~Allanach {\it et al.},
in {\it Proc. of the APS/DPF/DPB Summer Study on the Future of
Particle Physics (Snowmass 2001) } ed. N.~Graf,
Eur.\ Phys.\ J. {\bf C25} (2002) 113
[arXiv:hep-ph/0202233].

\bibitem{benchmark}
M.~Battaglia {\it et al.}, Eur. Phys. J.  {\bf C22} (2001) 535
[arXiv:hep-ph/0106204].





\end{thebibliography}
\end{document}